\documentclass[leqn,usenatbib]{mnras}
\usepackage[T1]{fontenc}
\usepackage{ae,aecompl}
\usepackage{graphicx}	% Including figure files
\usepackage{amsmath}	% Advanced maths commands
\usepackage{amssymb}	% Extra maths symbols
\title[Discordance of dipole asymmetries in large radio surveys with Cosmological Principle]{Discordance of dipole asymmetries seen in recent large radio surveys with the Cosmological Principle}
\author[A. K. Singal]{Ashok K. Singal\thanks{E-mail: ashokkumar.singal@gmail.com}\\
{Astronomy and Astrophysics Division, Physical Research Laboratory, 
Navrangpura, Ahmedabad - 380009, India}}
\date{Accepted XXX. Received YYY; in original form ZZZ}
\pubyear{2023}
\begin{document}
\label{firstpage}
\pagerange{\pageref{firstpage}--\pageref{lastpage}}
\maketitle

% Abstract of the paper
\begin{abstract}
In recent years, large radio surveys of Active Galactic Nuclei (AGNs), comprising millions of sources, have become available where one could investigate dipole asymmetries, assumedly arising due to a peculiar motion of the Solar system. Investigations of such dipoles have yielded in past much larger amplitudes than the cosmic microwave background (CMB) dipole, though their directions seem to lie close to the CMB dipole. Here we investigate dipole  asymmetries in two recent large radio surveys, Very Large Array Sky Survey (VLASS) containing 1.9 million sources, covering the sky north of $-40^\circ$ declination, and the Rapid ASKAP Continuum Survey (RACS) containing 2.1 million sources, covering the sky south of  $+30^\circ$ declination. We find  dipoles determined from the VLASS and RACS surveys to be significantly larger than the CMB dipole. Dipole directions from the VLASS and RACS data differ significantly from each other. Nevertheless, along with a number of other previously determined dipoles, including the CMB, they all appear to lie in a narrow sky region, which argues for the various dipoles to be related somehow. However, significant differences in their derived peculiar velocities, including that of the CMB, cannot be explained by a peculiar motion of the Solar system, which should necessarily be a single value. Instead, their discordant peculiar velocities may be indicating that different cosmic reference frames are moving relative to each other or that the matter distribution on cosmic scales is not homogeneous and isotropic, either scenario being in contravention of what expected from the Cosmological Principle (CP). 
\end{abstract}
%--------------------------
\section{INTRODUCTION}\label{S0}
According to the CP, on a sufficiently large scale the Universe is homogeneous and isotropic, and an observer stationary with respect to the comoving coordinates of the cosmic fluid, should find the distribution of distant AGNs
to be uniform over the sky. However, an observer moving with a velocity $v$ relative to the cosmic fluid will, as a combined effect of aberration and Doppler boosting, notice in the number counts as well as in the sky brightness a dipole anisotropy of an amplitude
%therefrom 
%to vary by a factor $\propto \delta^{2+x(1+\alpha)}$, where $\delta=1+(v/c)\cos\theta$, 
%is the Doppler factor, assuming it to be a non-relativistic motion as shown by all previous observations (Lineweaver et al. 1996; Hinshaw et al. 2009; Singal 2011,19a,b,21,22a,b; Rubart \& Schwarz 2013; Tiwari et al. 2015; Colin et al. 2017; Bengaly et al. 2018; Aghanim et al. 2020; Siewert et al. 2021; Secrest et al. 2021,22).
%%\cite{1,2,3,7,8,9,19,12,Si19a,Si19b,Se21,Sie21,Si21,Si22a,Si22b,Se22}.
%%\cite{6,7,4,10}.  
%This implies the number counts and the sky brightness will 
\begin{equation}
\label{eq:1}
{\cal D}=\left[2+{x(1+\alpha)}\right]\frac{v}{c}\;.
\end{equation}
Here $c$ is the velocity of light, $\alpha$ ($\approx 0.8$) is the spectral index, defined by $S \propto \nu^{-\alpha}$, and $x$ is the index of the integral source counts of extragalactic radio source population, which follows a power law, $N(>S)\propto S^{-x}$ ($x \sim 1$) (Ellis \& Baldwin 1984; Crawford 2009; Singal 2011,14). 

Within the CP, various cosmic reference frames, defined by the distant matter, should be coincident with the reference frame defined by the CMB, with no relative motion with respect to each other, then it is expected that the Solar system peculiar motion determined with respect to any  set of distant AGNs should be the same as with respect to the CMB. However, as was first pointed out by  
Singal (2011), the amplitude of the peculiar motion of the Solar system with respect to the reference frame defined by extragalactic radio sources seems to be about four times larger than   the value $370$ km s$^{-1}$, determined from the CMB reference frame (Lineweaver et al. 1996; Hinshaw et al. 2009; Aghanim et al. 2020). Surprisingly though, the direction of the velocity vector coincided, within statistical uncertainties, with the CMB dipole. Subsequent 
investigations have repeatedly shown that not only various cosmic reference frames seem to have relative motion with respect 
to the CMB reference frame, they do not seem to coincide among themselves, though the direction of motion has almost always turned out to be close to the CMB dipole   
(Rubart \& Schwarz 2013; Tiwari et al. 2015; Colin et al. 2017; Bengaly, Maartens \& Santos 2018; Singal 2019a,b,21,22a,b; Siewert, Rubart \& Schwarz 2021; Secrest et al. 2021,22).
A significant difference between dipole amplitudes would imply a relative motion between the corresponding cosmic reference frames or inhomogeneities and anisotropies on cosmic scales, either of which will be against the CP on which the whole modern cosmology is based upon. 
%This makes it imperative that investigations of radio dipoles, possibly, be made using many more independent datasets. 

On the other hand a recent investigation (Darling 2022) of the number counts as well as the sky brightness, using a combined data from two independent catalogues, the  
Very Large Array Sky Survey (VLASS) (Gordon et al. 2021) and the Rapid ASKAP Continuum Survey (RACS) (Hale et al. 2021) seems to have yielded a value for the radio dipole, however, in apparent agreement with the CMB dipole, both in direction and amplitude, which contradicts almost all earlier findings for such  dipoles (Singal 2011,19a,b,21,22a,b; Rubart \& Schwarz 2013; Tiwari et al. 2015; Colin et al. 2017; Bengaly et al. 2018; Siewert et al. 2021; Secrest et al. 2021).
 %\cite{7,8,9,19,12,Si19a,Si19b,Se21,Sie21,Si21,Si22a,Si22b}. 
As has been emphasized (Secrest et al. 2022), if the combined catalogue gives results in agreement with the CMB dipole, individual catalogues, which are sufficiently large, too should yield similar results.  Because the question involved here is of rather crucial relevance for cosmological studies, where at stake is the CP itself, we investigate here the radio dipoles in each of the VLASS as well as RACS samples, individually, and in sufficient details.

\section{VLASS AND RACS DatasetS}\label{S1}
The Very Large Array Sky Survey (VLASS) at 3 GHz (Lacy et al. 2020), carried out at Karl G. Jansky Very Large Array, covers the sky north of $-40^\circ$ declination. A catalogue containing  1.9 million sources has been derived from “quicklook” images of the Very Large Array Sky Survey (Gordon et al. 2021). An alternate version of the above catalogue from an independent reduction of the survey data is also available (Bruzewski et al. 2021), which, in order to distinguish, we call here VLASS-B. Both have been derived from “quicklook” images from the basic survey, however there are some minor differences between the catalogues. For instance, the version by Gordon et al. (2021) did not apply astrometric source position corrections as was done by Bruzewski et al. (2021), while the latter have not corrected for the apparent slight underestimation of flux densities of the order of $\sim 10\%$ as done by  Gordon et al. (2021).

The Rapid ASKAP Continuum Survey (RACS), covering the sky south of $+30^\circ$ declination at 887.5 MHz, carried out using the Australian Square Kilometre Array Pathfinder (ASKAP) contains 2.1 million sources
% at $>5$ mJy 
(Hale et al. 2021).

In our investigations, we are looking for a dipole in distribution of source positions in the sky where any gaps in the sky coverage can affect the dipole estimates. However, exclusion of sky-strips, which affect the forward and backward measurements 
identically, for example symmetric strips in diametrically opposite regions on the sky, to a first order do not have systematic effects on the inferred direction of the dipole (Ellis \& Baldwin 1984), apart from the errors becoming larger because of lesser data. 
Since the VLASS catalogue (Gordon et al. 2021) has a gap of sources for Dec $<-40^{\circ}$, we drop all sources with Dec $> 40 ^{\circ}$ as well to have equal and opposite gaps on two opposite 
sides.   
We also exclude all sources from our sample which lie in the Galactic plane ($|b|<10^{\circ}$). Similarly from the RACS catalogue (Hale et al. 2021), we dropped all sources with 
Dec $<-30^{\circ}$ as well as $|b|<10^{\circ}$.

\begin{figure}
\includegraphics[width=\columnwidth]{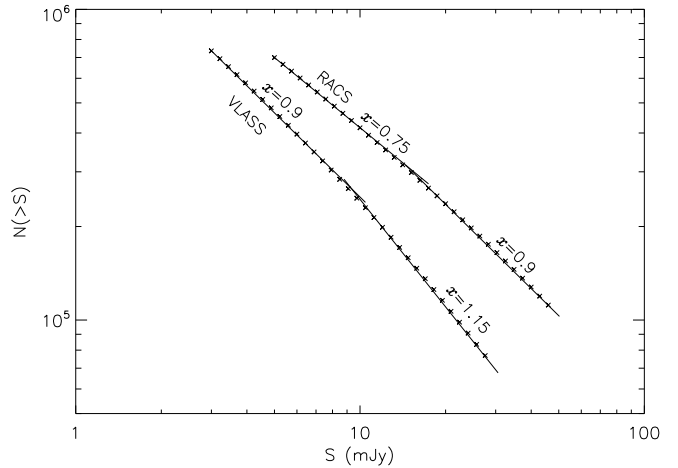}
\caption{A plot of the integrated source counts $N(>S)$ against $S$, for the VLASS and RACS  samples, showing the power law  behavior ($N(>S)\propto S^{-x}$) of the source counts. From  piece-wise straight line fits to data in different flux-density ranges in either sample, index $x$ appears to steepen for stronger sources, as shown by continuous lines with the best-fit index values written above.}
\label{F1}
\end{figure}

The integrated source counts, $N(>S)$ for different $S$ for the VLASS and RACS samples are shown in Fig.~\ref{F1}. The index $x$  in the power law relation, $N(>S)\propto S^{-x}$, estimated from the slope of the $\log N-\log S$ plot in Fig.~\ref{F1}, can be seen to steepen from low to high flux-density levels for both samples. From piece-wise straight line fits to the $\log N-\log S$ data, we find that $x$ steepens from  $0.9$ to $1.15$ around 10 mJy in the VLASS data, while in the RACS sample it steepens from  $0.75$  to $0.9$ around 15-20  mJy. These breaks in the index values seem to indicate intrinsic changes in the source count indices. 
%but these could also occur if there are some missing sources at low flux densities. To be on the safer side, 
We restrict our investigations to flux densities 10 mJy or above for the VLASS data and to flux densities 20 mJy or above for the RACS data and use the corresponding values of $x=1.15$ for the  VLASS sample and  $x=0.9$ for the VLASS sample in Eq.~(\ref{eq:1}), while deriving the peculiar velocities from the observed values of dipole amplitude, $\cal D$, in each case. As earlier measurements have shown the peculiar velocity estimates to be about 2 to 20 times higher than the CMB value of 370 km s$^{-1}$, for convenience of comparison we use a parameter $p$ for the amplitude of the peculiar velocity $v$, in units of the CMB value 370 km s$^{-1}$, so that $v=p \times 370$ km s$^{-1}$, with $p=0$ implying a nil peculiar velocity while $p=1$ implying the CMB value.
%--------------------------------------------
%--------------------------------------------
%--------------------------------------------

\section{Method and Procedure}\label{S2}
\subsection{Dipole vector determined from a sum of position vectors of the sources}\label{S2a}
We can determine the Solar system peculiar velocity from a dipole in the distribution of source positions in sky. First a unit vector $\bf{\hat{d}}$, in the dipole direction, is determined from the vector sum 
\begin{eqnarray}
\label{eq:1.5}
\bf{\hat{d}}=\frac{\sum_{i=1}^{N}\bf{\hat{r}}_{\rm i}}{|\sum_{i=1}^{N}\bf{\hat{r}}_{\rm i}|}\,,
\end{eqnarray}
where $\bf{\hat{r}}_{\rm i}$ is the angular position vector of $i$th source in sky and $N$ is the total number of sources in the sample used (Crawford 2009; Singal 2011). Then amplitude of the dipole is computed from  
\begin{eqnarray}
\label{eq:2}
{\cal D}&=&\frac{3}{2} \frac{\sum_{i=1}^{N} \bf{\hat{d}}\cdot\bf{\hat{r}}_{\rm i}}{\sum_{i=1}^{N} |\bf{\hat{d}}\cdot\bf{\hat{r}}_{\rm i}|}= \frac{3}{2}\frac{\sum_{i=1}^{N} \cos \theta_{\rm i}}{\sum_{i=1}^{N} |\cos \theta_{\rm i}|},
\end{eqnarray}
where $\theta_{\rm i}$ is the polar angle of the $i${th} source with respect to  $\bf{\hat{d}}$, the dipole direction determined in Eq.~(\ref{eq:1.5}). The statistical uncertainty in estimated $\cal D$ is $\sqrt {3/N}$ (Crawford 2009; Singal 2011).  

%To estimate statistical uncertainties in the sky position of the estimated dipole vector, we created an artificial radio sky with sources distributed at random positions in sky, but with a flux-density distribution as of the VLASS sample, so that the source counts remain unchanged. On this was superimposed a mock dipole, pointing in each simulation toward different direction in sky. From this mock data, the dipole was recovered and compared with the input dipole to estimate the difference in position. Five hundred such independent simulations were made to estimate the expected errors in the dipole co-ordinates. 

If in place of $\cos \theta_{\rm i}$ one uses  $S_{\rm i} \cos \theta_{\rm i}$ within summations in both numerator and denominator in Eq.~(\ref{eq:2}), where $S_{\rm i}$ is the observed flux density of the $i${th} source, then one begets dipole $\cal D$ from the sky brightness due to radio sources. Of course Eq.~(\ref{eq:1}) still gives the peculiar velocity from $\cal D$ thus computed. However, in Eq.~(6) of Darling (2022), following Rubart \& Schwarz (2013), an extra factor of $\delta^{1+\alpha}$ due to Doppler boosting in the integrated flux density per solid angle was included, which is erroneous. Actually in an {\em observed} flux-density range, which is chosen to be the same for all directions in sky, one multiplies $S$ with the number of sources visible to the observing instrument at that flux density level. For instance, the contribution to the sky brightness at an observed flux density $S$ comes from sources whose rest-frame flux density is $S/\delta^{1+\alpha}$. Thus the flux boosting of individual sources, pointed out in Darling (2022) for the formula already gets compensated for because of the fact that in the rest frame the sources were intrinsically weaker by a factor $\delta^{(1+\alpha)}$.
Of course, as a result, the number of sources at the flux density S alter   
by a factor $\delta^{x(1+\alpha)}$, while the Doppler boosting by the factor $\delta^{1+\alpha}$ is already accounted for in the altered number of sources in that observed flux-density bin for every direction in sky.
%As a result the number of sources 
%$(\mathrm{d}N/\mathrm{d}S) \Delta S$ seen in the moving observer's frame between $S$ and $S+ \Delta S$ will be the same as that lies between $S/\delta^{1+\alpha}$ and $(S+ \Delta S)/\delta^{1+\alpha}$ in the rest frame, and which is already accounted for in Eq.~(\ref{eq:1}) here. 
All this has been pointed out and discussed in detail already (Singal 2014), where it was shown that the correct formula to compute the peculiar velocity from the sky brightness is the same as in Eq.~(\ref{eq:1}) here, and which has been employed earlier for this purpose (Singal 2011).

Further, in the sky brightness method, a relatively small number of strong sources at high flux-density levels could introduce large statistical fluctuations (Singal 2011), therefore one ends up putting an upper limit on the flux density of sources in the sample to be employed. However, in the the number counts (Eq.~(\ref{eq:2})), unlike in the case of sky brightness, a small number of bright sources do not adversely affect the results, therefore, we have chosen to restrict our analysis here to number counts only.
%---------------------------------------------------
\begin{table*}
\begin{center}
\caption{\label{T1}}{Peculiar velocity vector from number counts for the VLASS dataset with $|{\delta}|<40^\circ$} %and $|b|>10^\circ$
\hskip4pc\vbox{\columnwidth=33pc
\begin{tabular}{ccccccccccccccc}
\hline\hline 
(1)&(2)&&(3)&&(4)&&(5)&&(6)&&(7)&&(8)\\
Flux-density & $N$ &&  RA && Dec && ${\cal D}$  &  & $p$&& ${\cal D}_{\rm h}$  &  & $p_{\rm h}$\\
 (mJy) && & ($^{\circ}$)& & ($^{\circ}$) &&  ($10^{-2}$) && ($370$ km s$^{-1}$)&&  ($10^{-2}$)&& ($370$ km s$^{-1}$) \\ \hline
$\geq 50$ & 36532 &&  $160\pm 21$ &&  $00\pm 27$ && $2.2\pm 1.0$ && $4.4\pm2.0$ && $1.9\pm 1.0$ && $3.8\pm2.1$ \\
$\geq 30$ & 69465 &&  $174\pm 15$ &&  $13\pm 25$ && $1.9\pm 0.7$ && $3.8\pm1.4$ && $2.2\pm 0.8$ && $4.4\pm1.5$ \\
$\geq 20$ & 112322 &&  $188\pm 14$ &&  $28\pm 23$ && $1.8\pm 0.6$ && $3.5\pm1.1$ && $2.5\pm 0.6$ && $5.0\pm1.2$ \\
$\geq 10$ & 240458 &&  $189\pm 12$ &&  $42\pm 22$ && $1.9\pm 0.4$ && $3.8\pm 0.8$ && $2.1\pm 0.4$ && $4.3\pm0.8$ \\
%$\geq 5$ & 470760 &&  $166\pm 09$ &&  $68\pm 08$ && $1.41\pm 0.30$ && $1.11\pm0.24$ && $1.97\pm 0.29$ && $1.56\pm0.23$ \\
%$\geq 3$ & 735449 &&  $110\pm 09$ &&  $76\pm 08$ && $2.15\pm 0.24$ && $1.69\pm0.19$ && $2.10\pm 0.23$ && $1.66\pm0.18$ \\
\hline
\end{tabular}
}
\end{center}
\end{table*}
%--------------------------------------------
\begin{table*}
\begin{center}
\caption{\label{T2}}{Peculiar velocity vector from number counts for the VLASS-B dataset with $|{\delta}|<40^\circ$} %and $|b|>10^\circ$
\hskip4pc\vbox{\columnwidth=33pc
\begin{tabular}{ccccccccccccccc}
\hline\hline 
(1)&(2)&&(3)&&(4)&&(5)&&(6)&&(7)&&(8)\\
 Flux-density & $N$ &&  RA && Dec && ${\cal D}$  &  & $p$&& ${\cal D}_{\rm h}$  &  & $p_{\rm h}$\\
 (mJy) && & ($^{\circ}$)& & ($^{\circ}$) &&  ($10^{-2}$) && ($370$ km s$^{-1}$)&&  ($10^{-2}$)&& ($370$ km s$^{-1}$) \\ \hline
$\geq 50$ & 35447 &&  $182\pm 20$ &&  $01\pm 27$ && $1.1\pm 1.0$ && $2.3\pm2.0$ && $1.7\pm 1.1$ && $3.4\pm2.1$ \\
$\geq 30$ & 68833 &&  $176\pm 16$ &&  $22\pm 25$ && $1.3\pm 0.7$ && $2.5\pm1.4$ && $1.9\pm 0.8$ && $3.8\pm1.5$ \\
$\geq 20$ & 112017 &&  $192\pm 15$ &&  $46\pm 24$ && $1.6\pm 0.6$ && $3.1\pm1.2$ && $1.9\pm 0.6$ && $3.9\pm1.2$ \\
$\geq 10$ & 237456 &&  $215\pm 15$ &&  $52\pm 23$ && $2.1\pm 0.4$ && $4.1\pm 0.8$ && $2.2\pm 0.4$ && $4.5\pm0.8$ \\
%\hline
\hline
\end{tabular}
}
\end{center}
\end{table*}
%---------------------------------------------------------

%----------------------------------------
\subsection{Hemisphere method with respect to the estimated dipole direction}\label{S2b} 
Dipole magnitude can be estimated in another way called the hemisphere method, which, unlike the vector method described in section~\ref{S2a}, does not directly yield the direction of the dipole. Since we do not know the dipole direction, we have to start with a certain assumed direction for the dipole and compute the magnitude of the dipole in that direction. It could, for example, be taken from the already known direction, like that of the CMB, or it be taken from the direction $\bf{\hat{d}}$ of the dipole as determined from the vector method. Then the sky is divided in two hemispheres, $H1$ and $H2$, with $H1$  centred on the assumed direction of the dipole and $H2$ centred on the opposite direction. If $N_1$ is the number of sources found in $H1$  and  $N_2$ is that found in $H2$, then the dipole ${\cal D}$ could be determined from the observed fractional excess of $N_1$ over $N_2$ as 
\begin{equation}
\label{eq:3}
{\cal D}=2\frac {\delta N}{N}=2\frac {N_1 - N_2} {N_1 + N_2}.
\end{equation}
while the error in $\cal D$ in the hemisphere method is $2/\sqrt {N}$ (Singal 2019a).

However, we may not want to get biased by an already known direction like that from the CMB, and might like to determine the direction of the dipole independent of the vector method as well. In that case, we can employ a `brute force' method (Singal 2019b).
We divide the sky into small pixels of angular area, say $\Delta\theta\times \Delta\theta$, creating a grid of $n$ cells covering the whole sky area of $4\pi$ sr, with minimal overlaps. 
Then one by one, taking the trial pole direction to be the centre of each of these $n$ pixels, and accordingly counting sources in our sample in the two hemispheres with respect to that trial pole direction, we compute the dipole amplitudes ($p$), using Eqs.~(\ref{eq:1}) and (\ref{eq:3}). Thus for each one of these $n$ pixels, we have RA, Dec, and a peculiar velocity value $p$. However, this $p$ value may only be  a projection of the true peculiar velocity along the corresponding RA and Dec. Therefore, we can expect a peak along the real dipole direction, along with a $\cos\psi$ dependence in the $p$ values, determined for the other $n-1$ grid points around it.

The~location of the peak value for the dipole amplitude, in principle, should yield the true direction of the dipole. However, because of statistical fluctuations in individual values,
it may not always be possible to zero down on a single unique peak for the true dipole direction. Nevertheless, we can refine the procedure for determining the pole direction by making use of the expected $\cos\psi$ dependence of $p$ for grid points at polar angle $\psi$ from the true pole. 

We have written a COSFIT routine which, for each one of the $n$ sky positions, makes 3-d cos fits to the $p$ values of surrounding $n-1$ pixels around it, and determines the sky position of the pixel that yields the highest value, which then would be the optimum direction for the peculiar motion. One could also evaluate the $\chi^2$ value from the residuals to each of these $n$ 3-d cos fits to the expected $\cos\psi$ dependence, its minimum should also occur close to the true direction of the peculiar motion. 
%%--------------------------------------------
\begin{figure}
\includegraphics[width=\linewidth]{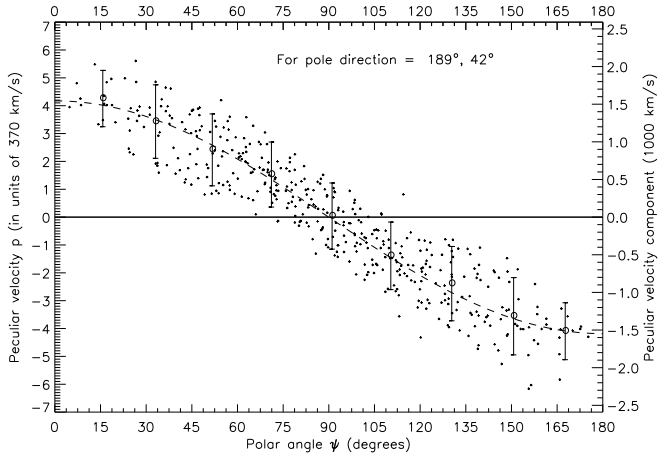}
\caption{Variation of the peculiar velocity component $p$ (in units of CMB value 370 km s$^{-1}$), computed for various polar angles %(a) 
with respect to the dipole direction, RA$=189^{\circ}$, Dec$=42^{\circ}$, derived by the dipole vector method for the $S\geq 10$ mJy sub-sample of the VLASS data (Table~\ref{T1}). 
The corresponding peculiar velocity values of the Solar system in units of $10^3$ km s$^{-1}$ are shown on the right hand vertical scale.
Plotted circles (o) with error bars show 
values for bin averages of the peculiar velocity components, obtained for various $20^{\circ}$ wide slices of the sky in polar angle,  
while the dashed line 
shows a least square fit of $\cos \psi$ to the bin average values.\label{F2}
}
\end{figure}
%%--------------------------------------------
%--------------------------------------------
\section{Results and Discussion}\label{S4}
The results for the dipole, determined using  the vector method (section~\ref{S2a}) from the anisotropy in number counts in the VLASS sub-samples  at four different flux-density levels, are presented in Table~\ref{T1}, where column (1) gives the flux-density limit of the sub-sample, column (2) gives the number of sources in the sub-sample, columns (3) and (4) list the direction of the dipole in terms of right ascension and declination, derived from the vector dipole method applied to that sub-sample, column (5) gives $\cal D$, the~dipole magnitude computed from Eq.~(\ref{eq:2}), and column (6) lists  $p$, the peculiar speed estimated from $\cal D$ using Eq.~(\ref{eq:1}). 
Columns (7) and (8) list dipole ${\cal D}_{\rm h}$, and peculiar speed, $p_{\rm h}$, determined from the hemisphere method (section~\ref{S2b}), for the direction given in columns (3) and (4) for the corresponding sub-sample.
From Table~\ref{T1} we find the peculiar speed from the VLASS data in various flux-density bins to be about four times the peculiar speed estimated from the CMB dipole.
The peculiar velocity $v$ can be calculated from $p$ as $v=p \times 370$ km s$^{-1}$.

From Table~\ref{T1} we see a trend that the direction of the dipole from VLASS data seems to shift northward with lower flux-density levels. In fact the direction of the dipole at $\geq 30$ mJy levels might appear to be in agreement with the CMB dipole (\mbox{RA$=168^{\circ}$}, Dec$=-7^{\circ}$), but as we go to the lower flux-density levels ($\geq 10$ mJy) the direction of the determined dipole shifts significantly away from the CMB dipole, especially in declination.
 
In order to ascertain whether this trend is genuinely present in the VLASS data, we have also determined the dipoles in the VLASS-B data (Bruzewski et al. 2021) and the results are presented in Table~\ref{T2}. Although entries in Table~\ref{T1} and Table~\ref{T2} may seem to differ in details, the overall results appear to be in agreement. A similar trend of a shift in the dipole position in declination with decreasing flux levels is seen in both Table~\ref{T1} and Table~\ref{T2}.

To determine the dipole direction in sky from the hemisphere method (section~\ref{S2b}) using the brute force technique (Singal 2019b), we first divided the sky into $10^\circ \times 10^\circ$ pixels with minimal overlap, thereby creating a grid of $422$ cells covering the whole sky. 
Then taking the dipole direction to be the centre of each of these 422 cells, in turn, counted the number of sources in corresponding hemispheres $H1$ and $H2$ for a sample chosen from say,  $S\geq 10$ mJy (Table~\ref{T1}), and using Eq.~(\ref{eq:3}) computed the dipole magnitude $p_{\rm i}$  for $i=1$ to $422$. Actually this yields only a projection of the peculiar velocity in the direction of $i$th pixel, which should have a $\cos\psi_{\rm i}$ dependence where $\psi_{\rm i}$ is a polar angle of $i$th pixel with respect to the actual pole. 
%%--------------------------------------------
\begin{figure}
\includegraphics[width=\columnwidth]{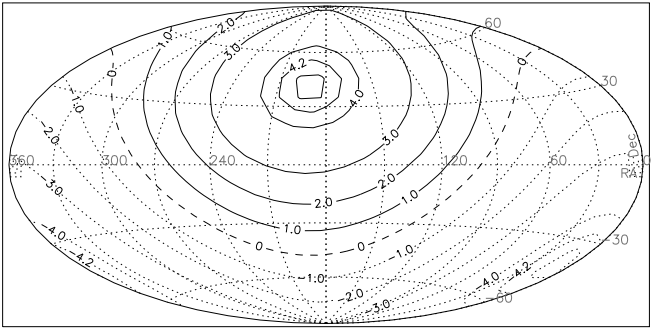}
\caption{A contour plot of 3-d  cos fits  made to the dipole amplitudes, $p$, estimated for various directions across the sky for the $S\geq 10$ mJy sub-sample of the VLASS data, showing a unique unambiguous peak indicating the optimum direction of the dipole. The~horizontal and vertical axes denote RA, from $0^\circ$ to $360^\circ$, and Dec from $-90^\circ$ to $90^\circ$.
The positions of the peak yields the direction of the observer's peculiar velocity at RA = 189$^{\circ}$, and~Dec = 42$^{\circ}$, consistent with that obtained from the sum of position vectors for this sub-sample (Table~\ref{T1}). 
\label{F3}
}
\end{figure}
%%--------------------------------------------

In order to test this $\cos\psi$ dependence, we have plotted in Fig.~(\ref{F2}) the peculiar velocity components $p_{\rm i}$  for various $n$ points, as a scatter plot for different $\psi_{\rm i}$ values, measured with respect to the direction,  RA$=189^{\circ}$, Dec$=42^{\circ}$, derived from dipole vector method, (Table~\ref{T1},  $S\geq 10$ mJy). 
%The scatter plot and their various bin-average values clearly show the expected $\cos\psi$ behaviour with a maximum value of $p=4.3$ (Fig.~\ref{F2}). 
We also computed bin averages of peculiar velocity $p$ in $20^{\circ}$ wide slices of the sky by divided the sky into bins of $20^\circ$ width in polar angle about the above pole position, viz. RA$=189^{\circ}$, Dec$=42^{\circ}$. A least square fit of $\cos \psi$ to the bin average values (Fig.~(\ref{F2})) shows that the computed $p$ values for various pixels at polar angles ($\psi$) do follow a systematic $\cos\psi$ dependence. 
%%--------------------------------------------
\begin{figure}
\includegraphics[width=\columnwidth]{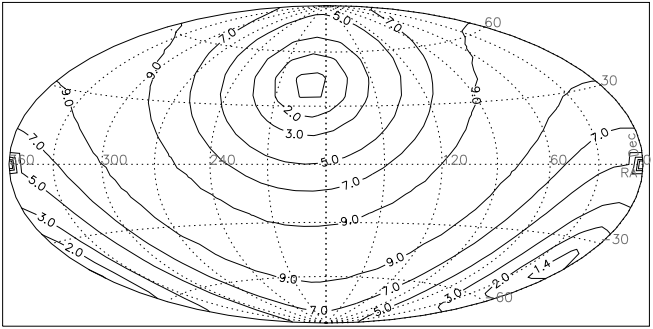}
\caption{Reduced Chi-squared ($\chi^2_\nu$) values for the 3-d  cos fits, made to the dipole amplitudes, $p$, estimated for various directions across the sky for the $S\geq 10$ mJy sub-sample of the VLASS data, shows a minimum value of 1.2, quite close to the ideal minimum value of unity, at exactly the same sky position, RA = 189$^{\circ}$, and~Dec = 42$^{\circ}$, as the peak in Fig.~(\ref{F3}). \label{F4}
}
\end{figure}
%--------------------------------------------

We now made a 3-d $\cos\psi$ fit for each of the $n=422$ positions for the remaining $n-1$ $p$ values, and also computed the chi-square value for each of these $n$ fits. 
This COSFIT routine resulted in a clear unique peak, indicating the optimum direction of the dipole (Fig.~(\ref{F3})). A reduced Chi-squared ($\chi^2_\nu$) values for the 3-d  cos fits  made to the dipole amplitudes estimated for various trial dipole directions across the sky, shows a minimum value of 1.2 (Fig.~(\ref{F4})), quite close to the ideal minimum value of unity (Bevington \& Robinson 2003), at~the same sky position as the peak in (Fig.~(\ref{F3})). Thence we infer the direction of the observer's peculiar velocity as RA = 189$^{\circ}$, and~Dec = 42$^{\circ}$, which agrees very well with the corresponding value (RA = 189$^{\circ}$, Dec = 42$^{\circ}$), derived directly from the dipole vector method (Table~1, $S_{\rm 3GHz}\geq 10$ mJy).
We also tried finer grids with $5^\circ \times 5^\circ$ bins with $1668$ cells and even a grid with $2^\circ \times 2^\circ$ bins with $10360$ cells, but it made no perceptible difference in the results. 
%--------------------------------------------
\begin{figure*}
\includegraphics[width=\linewidth]{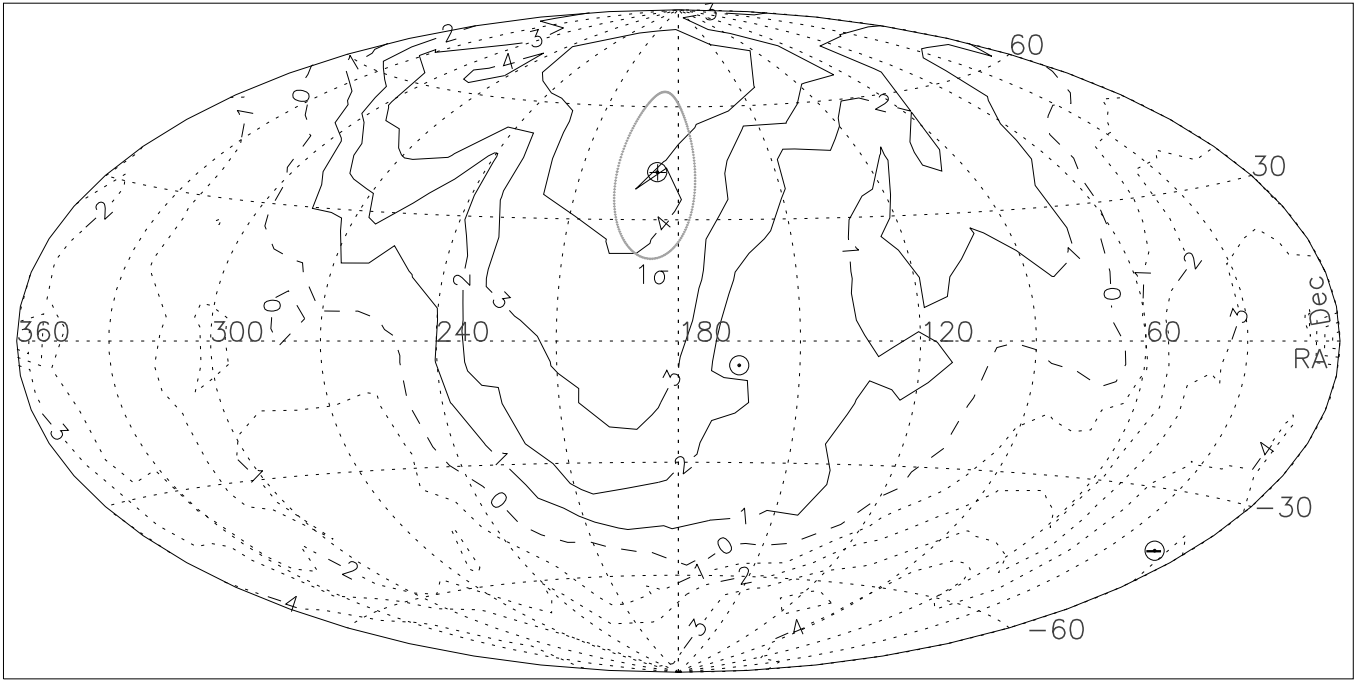}
\caption{A contour map of the dipole amplitudes, in the Hammer--Aitoff equal-area projection on the sky, estimated for various directions in the sky, for the $S\geq 10$ mJy sub-sample of the VLASS data. The~horizontal and vertical axes denote RA, from $0^\circ$ to $360^\circ$, and Dec from $-90^\circ$ to $90^\circ$. The~true pole direction is expected to be closer to the higher contours values, shown by continuous lines, while the true antipole should lie closer to the lower contour values, shown by dotted lines. The~dashed curve represents the zero amplitude of the dipole. 
The symbol $\oplus$ indicates the best-fit pole position for the VLASS sample, derived using our COSFIT routine to minimize $\chi^2$ (see the text), 
while the symbol $\ominus$ indicates the corresponding antipole position. 
%Two gray-colour error ellipses  around the best-fit position $\oplus$ represent the $1\sigma$ ($68.3\%$) and $2\sigma$ ($95.5\%$) confidence limits.
The gray-colour error ellipse  around the best-fit position $\oplus$ represents the $1\sigma$ ($68.3\%$) confidence limits. The symbol $\odot$ indicates the CMB pole position. 
\label{F5}
} 
\end{figure*}
%--------------------------------------------

%---------------------------------------------------
\begin{table*}
\begin{center}
\caption{\label{T3}}{Peculiar velocity vector from number counts for the RACS dataset with $|{\delta}|<30^\circ$} and $|b|>10^\circ$
\hskip4pc\vbox{\columnwidth=33pc
\begin{tabular}{ccccccccccccccc}
\hline\hline 
(1)&(2)&&(3)&&(4)&&(5)&&(6)&&(7)&&(8)\\
 Flux-density & $N$ &&  RA && Dec && ${\cal D}$  &  & $p$&& ${\cal D}_{\rm h}$  &  & $p_{\rm h}$\\
 (mJy) && & ($^{\circ}$)& & ($^{\circ}$) &&  ($10^{-2}$) && ($370$ km s$^{-1}$)&&  ($10^{-2}$)&& ($370$ km s$^{-1}$) \\ \hline
$\geq 100$ & 49711 &&  $200\pm 25$ &&  $-39\pm 31$ && $4.6\pm 0.9$ && $10.2\pm2.0$ && $4.5\pm 0.9$ && $10.0\pm2.0$ \\
%$\geq 70$ & 73142 &&  $201\pm 19$ &&  $-38\pm 30$ && $4.27\pm 0.72$ && $9.3\pm1.5$ && $4.11\pm 0.74$ && $9.0\pm1.6$ \\
$\geq 50$ & 102837 &&  $193\pm 17$ &&  $-41\pm 30$ && $3.9\pm 0.6$ && $8.7\pm1.4$ && $3.7\pm 0.6$ && $8.2\pm1.4$ \\
$\geq 30$ & 166423 &&  $197\pm 14$ &&  $-42\pm 28$ && $3.9\pm 0.5$ && $8.7\pm1.1$ && $3.3\pm0.5$ && $7.3\pm 1.1$\\
$\geq 20$ & 236887 &&  $197\pm 13$ &&  $-43\pm 23$ && $4.0\pm 0.4$ && $8.9\pm0.9$ && $3.3\pm 0.4$ && $7.3\pm0.9$ \\
\hline
\end{tabular}
}
\end{center}
\end{table*}
%---------------------------------------------------

The dipole amplitude distribution across the sky is shown in a contour map for the VLASS data for the $S\geq 10$ mJy sub-sample (Fig.~\ref{F5}). 
A broad plateau showing maxima in $p$, towards certain sky directions is clearly seen, however, from that it is not possible to zero down on a single unique peak for the true dipole direction. Therefore we have indicated in Fig.~\ref{F5} the 
optimum direction for the VLASS dipole, determined from our COSFIT routine to minimize $\chi^2$, by the symbol $\oplus$. The symbol $\odot$ indicates the CMB pole position on the map. 
%%--------------------------------------------

\begin{figure*}
\includegraphics[width=\linewidth]{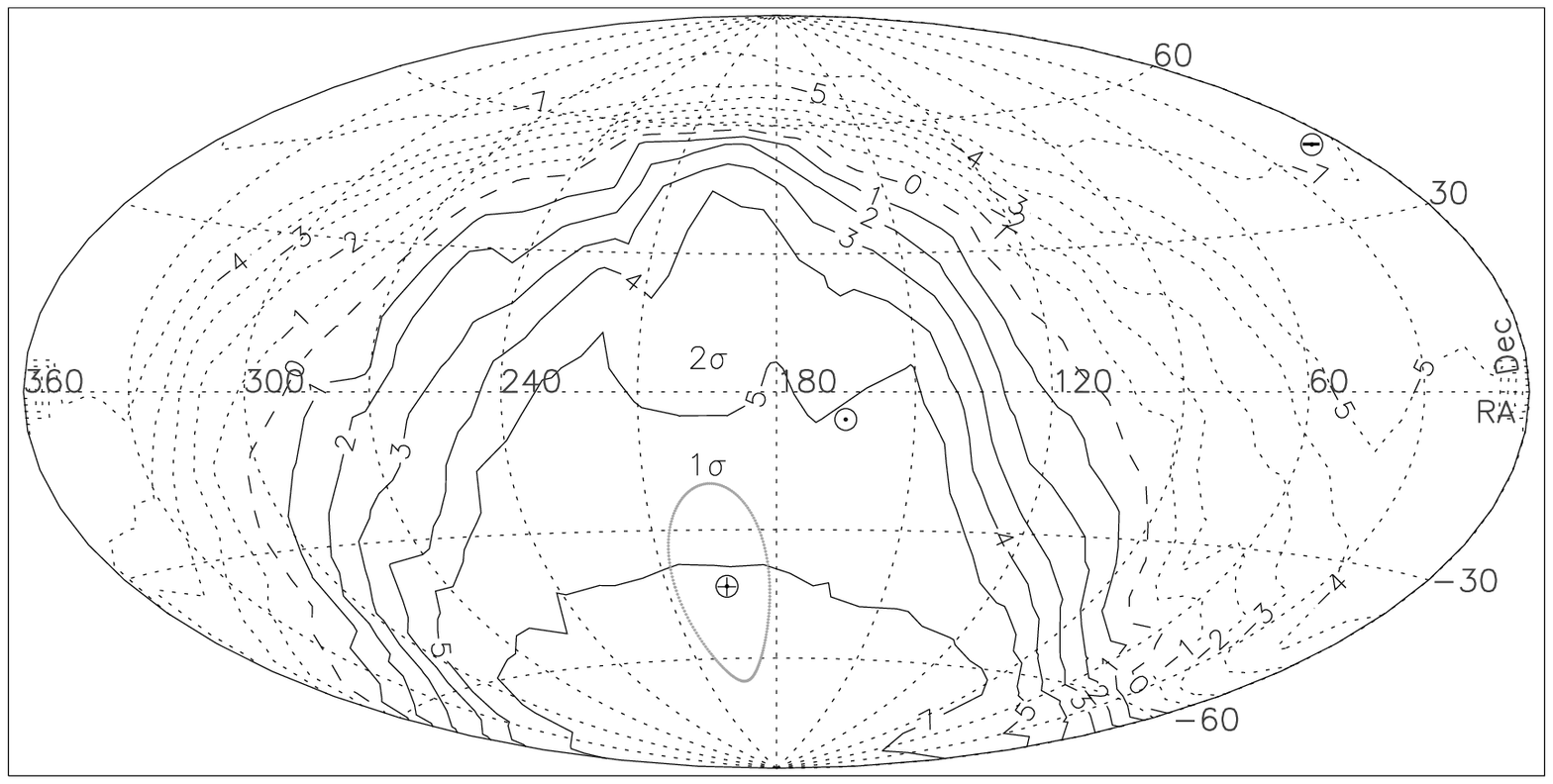}
\caption{A contour map of the dipole amplitudes for the $S\geq 20$ mJy sub-sample of the RACS data, estimated for various directions in the sky. The~horizontal and vertical axes denote RA and Dec in degrees. The~true pole direction is expected to be closer to the higher contours values, shown by continuous lines, while the true antipole should lie closer to the lower contour values, shown by dotted lines. The~dashed curve represents the zero amplitude of the dipole. 
The symbol $\oplus$ indicates the best-fit pole position for the RACS sample, derived using our COSFIT routine to minimize $\chi^2$, 
while the symbol $\ominus$ indicates the corresponding antipole position. 
The gray-colour error ellipse  around the best-fit position $\oplus$ represents the $1\sigma$ ($68.3\%$) confidence limits. The symbol $\odot$ indicates the CMB pole position. 
 \label{F6}
} 
\end{figure*}

A linear estimator like Eq.~(\ref{eq:1.5}) can introduce a bias in the dipole direction because of the incomplete sky coverage. However, one can avoid such a directional bias by introducing a symmetric cut in declination (Rubart \& Schwarz 2013), such that both polar caps are missing, a strategy first used by Singal (2011) and also employed here. This was tested from 500 simulations we made, using a different mock dipole vector in sky for each simulation. Moreover, we also estimated the dipole direction from the 3d COSFITs made to the dipole amplitudes for various directions in sky, computed from the hemisphere method (Section~\ref{S2b}). The fact that the dipole direction, both from the peak amplitude of the 3d COSFIT as well as from the $\chi^2_{red}$ agreed with that from the linear estimator, argues well for the dipole direction value being free of any bias due to our methods employed.

The results for the dipole, determined from the anisotropy in number counts in the RACS sample  at four different flux-density levels, are presented in Table~\ref{T3},  
where we see that the direction of the dipole is almost independent of the chosen flux-density levels, but is significantly away from the CMB dipole (\mbox{RA$=168^{\circ}$}, Dec$=-7^{\circ}$). 
The dipole amplitude distribution across the sky for the RACS data is shown in a contour map for the $S\geq 20$ mJy sub-sample (Fig.~\ref{F6}). Again, a broad plateau showing maxima in $p$, towards certain sky directions is clearly seen, however, from that it is not possible to zero down on a single unique peak for the true dipole direction. Therefore we used our COSFIT routine to find the 
optimum direction for the peculiar motion, which is indicated in Fig.~\ref{F6} by the symbol $\oplus$. The symbol $\odot$ indicates the CMB pole position on the map.

A major effort was put into testing the procedure and estimation of uncertainties, especially in the sky position estimates of the dipoles. For that we used Monte--Carlo simulations to create an artificial radio sky with similar number densities of sources as in each of 
the two catalogues, by randomly assigning sky positions, for the observed flux-density values  in each sub-sample, so that overall source counts remain unchanged. 
On these we superimposed Doppler boosting and aberration effects of an assumed Solar system peculiar motion, choosing in each simulation a different direction in sky for the assumed velocity vector. 
This mock catalogue was then used to retrieve the velocity vector under the conditions of zone of avoidance in the galactic plane ($|b|<10^\circ$) or declination restrictions ($|\delta| < 40 ^\circ$ or $|\delta| < 30 ^\circ$, as the case may be) similar as in our actual VLASS or  RACS samples and compared with the input velocity vector in that particular realization. This not only validated our 
procedure as well as the computer routine, but also helped us make an estimate of errors in the dipole co-ordinates from 500 simulations we made in either case, using a different mock dipole vector in sky for each simulation.

%%--------------------------------------------
\begin{figure*}
\includegraphics[width=\linewidth]{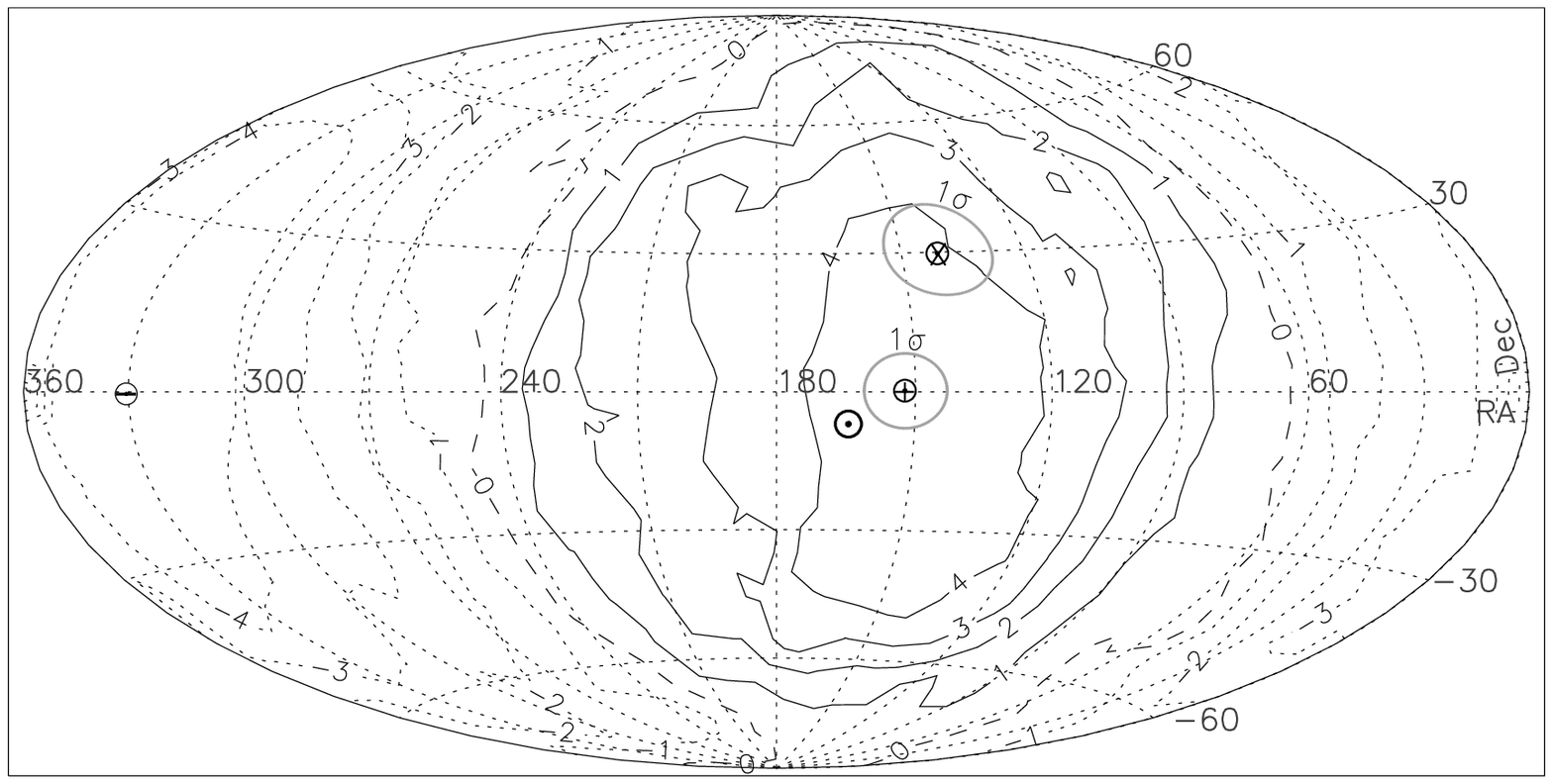}
\caption{A contour map of the dipole amplitudes for the NVSS data, estimated for various directions in the sky. The~horizontal and vertical axes denote RA and Dec in degrees. The~true pole direction is expected to be closer to the higher contours values, shown by continuous lines, while the true antipole should lie closer to the lower contour values, shown by dotted lines. The~dashed curve represents the zero amplitude of the dipole. 
The symbol $\oplus$ indicates the best-fit pole position for the NVSS sample, derived using our   COSFIT routine to minimize $\chi^2$, 
while the symbol $\ominus$ indicates the corresponding antipole position. 
The symbol $\otimes$ indicates another estimate of the dipole position, derived from an alternate method (see text). 
The gray-colour error ellipses around $\oplus$ and $\otimes$ represent the $1\sigma$ ($68.3\%$) confidence limits about the corresponding pole positions.  
%The symbol $\odot$ indicates the CMB pole position. 
\label{F7}
} 
\end{figure*}
%%%--------------------------------------------
\begin{figure*}
\includegraphics[width=\linewidth]{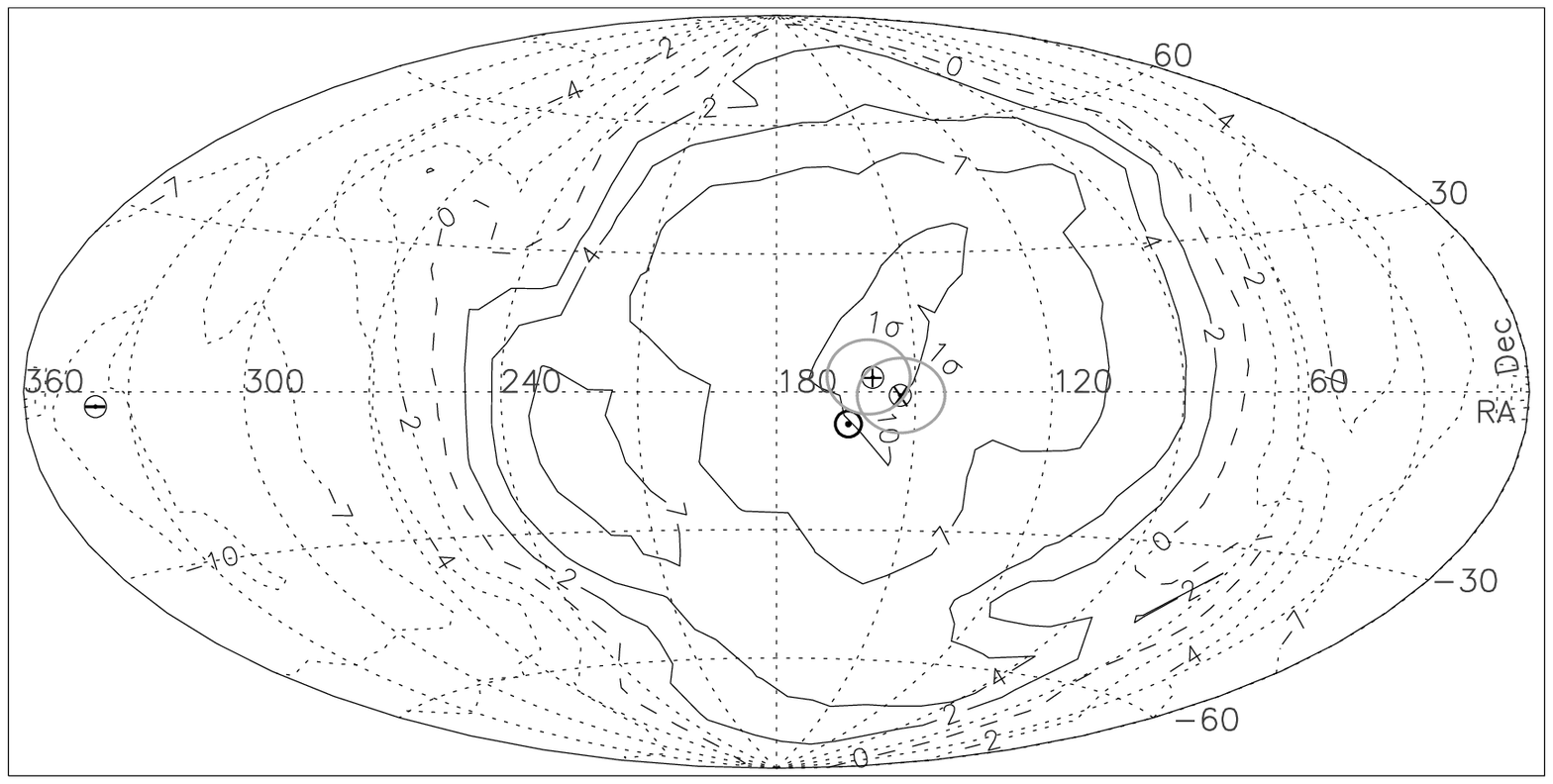}
\caption{A contour map of the dipole amplitudes for the TGSS data, estimated for various directions in the sky. The~horizontal and vertical axes denote RA and Dec in degrees. The~true pole direction is expected to be closer to the higher contours values, shown by continuous lines, while the true antipole should lie closer to the lower contour values, shown by dotted lines. The~dashed curve represents the zero amplitude of the dipole. 
The symbol $\oplus$ indicates the best-fit pole position for the TGSS sample, derived using our COSFIT routine to minimize $\chi^2$, 
while the symbol $\ominus$ indicates the corresponding antipole position. 
The symbol $\otimes$ indicates another estimate of the dipole position, derived from an alternate method  (see text). The gray-colour error ellipses around $\oplus$ and $\otimes$ represent the $1\sigma$ ($68.3\%$) confidence limits about the corresponding pole position.  
The symbol $\odot$ indicates the CMB pole position.
\label{F8}
} 
\end{figure*}
%%--------------------------------------------

If we now compare the dipole determined from number counts for the VLASS dataset (Table~\ref{T1} and Fig.~\ref{F5}) with that determined from the RACS dataset (Table~\ref{T3} and Fig.~\ref{F6}), we find that not only the amplitudes of the dipoles but also the directions of the dipoles from both these datasets differ significantly from each other as well as from the CMB dipole. The sky coverage so much overlaps and most sources must be common in the VLASS and RACS surveys, so one should expect the dipole directions to be similar in the two surveys.
Assuming there are no declination-dependent calibration systematics in either survey, a  statistically significant difference in the estimates of the two dipoles is puzzling.
 It is further intriguing that RACS data yields the pole positions almost independent of the flux-density levels, but in VLASS data there seems to be a systematic shift to higher declinations with a decreasing flux-density level. 
%--------------------------------------------------

The right ascensions of the VLASS dipole and  the RACS dipole do seem to agree reasonably well, which might indicate to some extent that the dipoles are genuine. 
Could there be a declination-dependent asymmetry in both surveys because
they are less sensitive north (RACS) or south (VLASS) of the equator? Each survey might be more  sensitive in the hemisphere of the physical location of the instrument. Thus one might expect the RACS dipole to point south of the equator and the VLASS dipole to point north of the equator, which is exactly what is seen. At the same time a similar argument does not seem to work for the dipole direction from the NVSS data which was collected using the same instrument, though in a different configuration, but having the same declination coverage of sky, as the VLASS data (Singal 2011). 

Errors in dipole positions, especially in declination, are larger because the estimated dipoles are lying outside the data coverage area limit ($\pm 40^\circ$ for the VLASS and $\pm 30^\circ$ for the RACS), and due to which the dipole declination estimates are less constrained.
%--------------------------------------------------

A combined set of data from the VLASS and RACS catalogues  
seems to have yielded a value for the radio dipole in apparent agreement with the CMB dipole, both in direction and amplitude (Darling 2022). Actually, in the case of VLASS and RACS, individual dipoles (Fig.~\ref{F5} and \ref{F6}) seem to be lying in the northern and southern hemispheres, almost equally away from the celestial equator. When combined, the asymmetric distributions of number density of sources in opposite hemispheres in the VLASS and RACS samples partly cancel these asymmetries and the inferred dipole of the combined data gives a much reduced value for the dipole amplitude and the direction somewhere in the middle of the two individual dipoles, which happens in this case to be close to the CMB dipole, and such  apparently has been verified (Secrest et al. 2022). As has been emphasized earlier (Singal 2022a; Secrest et al. 2022), in combining data from two independent catalogues to determine a dipole from number count asymmetries, there is always a possibility of getting results which might not truly represent an actual dipole in sky. 
In fact, a combination of two catalogues, even without any genuine dipoles or asymmetries present in their number densities, but having slightly different source number densities and covering partly different regions of sky, could yield an artificial dipole vector. The source number densities could be different due to slight mismatch in calibrations or the catalogues could be at different frequency bands and a slight uncertainty in spectral information might result in slightly different number densities of sources. 
After all inference of a dipole like the CMB dipole needs {\em overall} number densities  to differ only by one part in $10^3$ in the two separate catalogues.
Thus such catalogues when combined could result in an artificial dipole because of such different number densities in two different coverage directions of the two catalogues. On the other hand, in such a combination of catalogues there might even be a cancellation of individual dipoles, which may otherwise be actually present in the two separate catalogues, as seems to be the case in VLASS and RACS catalogues.
However, Wagenveld, Klöckner \& Schwarz (2023) seem to have successfully applied Bayesian estimators to determine the cosmic radio dipole from a combination of two separate catalogues.

Two other large radio surveys, NVSS and TGSS, where dipoles have been determined in past (Singal 2011; Bengaly et al. 2018; Singal 2019a), have yielded dipoles with similar larger amplitudes (by a factor of 4 to 10) than the CMB, though directions derived using dipole vector method were found to be consistent with the CMB dipole. Here we take a fresh look at the NVSS and TGSS dipoles by determining their directions using the brute force method, by minimizing $\chi^2$ in the 3-d $\cos \psi$ fitting using our COSFIT routine. We show in Fig.~\ref{F7} a contour map of dipole components in different sky directions, determined for the NVSS sample with $S\geq 20$ mJy (Singal 2011). The best-fit pole position for the NVSS, derived from a minimum of the reduced Chi-squared value ($\chi^2_\nu \stackrel{<}{_{\sim}} 1$), employing our COSFIT routine, is within $\sim 1.5\sigma$ of the CMB pole position and lies inside a contour corresponding to a peculiar velocity $p\sim 4$. Also indicated in Fig.~\ref{F7} is another estimate of the NVSS dipole position, derived from an alternate method (Secrest et al. 2022), which also lies within the highest contour in Fig.~\ref{F7}, however, the reduced Chi-squared values from  the 3-d cos fits, made to the dipole amplitudes estimated around this particular direction, shows $\chi^2_\nu \approx 3.0$, much above the minimum value. It should be noted that the direction for our best-fit pole position comes not from the dipole amplitude values on or near the highest contour, in the local neighborhood of the peak alone. Instead this optimum direction is derived by giving due weight to the dipole components at far off points in the sky as well, and an optimum direction for the dipole is obtained by minimizing the $\chi^2$ by fitting $\cos\psi$ through our COSFIT routine to the dipole amplitudes of all points on the sky.  

In Fig.~\ref{F8} we have shown a contour map of dipole components in different sky directions, determined for the TGSS data for the $S\geq 100$ mJy sub-sample (Singal 2019a). The best-fit pole position for the TGSS sample, derived from a minimum of the reduced Chi-squared value ($\chi^2_\nu \stackrel{<}{_{\sim}} 1$), employing our COSFIT routine, is at $\sim 1\sigma$ of the CMB pole position. Also shown is another independent determination of the dipole from the TGSS data (Bengaly et al. 2018). The dipole from TGSS data seems to lie within a contour of rather high amplitude, $p \stackrel{>}{_{\sim}} 10$, consistent with earlier determinations (Bengaly et al. 2018; Singal 2019a).
%\cite{12,Si19a}.

It has been pointed out categorically that the rather high amplitude of dipole in the TGSS data could be a result of calibration problems (Secrest et al. 2022). Incidentally, the pole of the TGSS data turning out to be so close to CMB dipole (Fig.~\ref{F8}), in spite of the calibration errors, if any, cannot be just fortuitous and argues for the TGSS dipole to be reasonably a genuine one. Only in a rather contrived scenario would one expect such a thing to have occurred  otherwise. In fact, during a comparison of the flux-density distributions among common sources in the NVSS and TGSS, any mismatches were attributed by Secrest et al. (2022) to calibration errors in the TGSS data alone, which may or may not be fully justified. By applying calibration corrections, the number density of sources in the TGSS sample were apparently adjusted by Secrest et al. (2022) to match those of NVSS data, no wonder the dipole amplitude of  the `corrected' TGSS sample turned out to be similar as of the NVSS dipole.
Further, thus estimated calibration errors as a function of sky positions by Secrest et al. (2022), when transformed from galactic to equatorial coordinates, can be seen, at least to a first order, to be mainly  
declination dependent. Now the pole of the TGSS dipole turns out to be almost at the equator, with the antipole direction $180^\circ$ away, again on or very near the equator (Fig.~\ref{F8}). In such a case any declination-dependent calibration errors will affect the number counts in hemispheres centred on the pole and the antipole almost in the same way and thus not influence the amplitude of the dipole adversely. Whatever else might be the reason for the high amplitude of the TGSS dipole, it does not seem that the sort of calibration errors, as have been pointed out by Secrest et al. (2022), could mainly be the reason for it.

Similar apparent anomalies in dipole amplitudes have been seen in sources selected from other than the radio surveys as well. For instance, the AGNs picked from the Wide-field Infrared Survey Explorer (WISE) catalogue (Wright et~al. 2010; Mainzer~et~al.  2014) too have yielded dipoles much larger than the CMB dipole (Singal 2021; Secrest et al. 2022). 
A redshift dipole along the CMB dipole direction in a~homogeneously selected DR12Q sample of quasars was seen which, interpreted in terms of the Solar system peculiar motion, gave a velocity $\sim6.5$ times the CMB dipole in a direction directly opposite to, but~nonetheless parallel to, the~CMB dipole (Singal 2019b). 
Also a peculiar motion of the observer can introduce a dipole in the $m-z$ Hubble plot by affecting the observed redshifts ($z$) as well as magnitudes ($m$) of the sources adversely in two opposite hemispheres. This has been exploited to estimate peculiar velocities from the Hubble diagrams of Supernovae (SNe) Ia and of quasars with spectroscopic redshifts (Singal 2022a,b). The estimated dipole amplitude from SNe Ia has turned out to be $\sim 4$ times larger than the CMB dipole, though in the same direction. On the other hand Horstmann, Pietschke \& Schwarz (2022) found the Solar system peculiar motion to be somewhat lower than that inferred from the CMB dipole, but in the same direction, while Sorrenti, Durrer \& Kunz (2023) found the amplitude
roughly in agreement with the CMB dipole, though the direction differs at very high significance.  It should, however, be noted that singal (2022b)  had used the JLA sample (Betoule et al. 2014) while Horstmann et al. (2022) and Sorrenti et al. (2023) used mostly the 
Pantheon sample (Jones et al. 2018; Scolnic et al. 2018). The difference in results could be due to the difference in the two samples as it was shown in Singal (2022b) that the two samples match at low redshifts ($z\sim 0.06$), however, at higher redshifts they progressively depart in magnitude ($m_{\rm B}$), with the Pantheon sample being systematically brighter than the JLA sample at $z\sim 0.6$ by $\Delta m_{\rm B}\approx -0.5$.

An assertion has been made (Dalang \& Bonvin 2022) that the differences in dipole amplitudes from the CMB dipole is due to a problem in Eq.~(\ref{eq:1}), where luminosity evolution of the AGN population has not been taken into account. In fact, it has been claimed that if effect of such redshift-dependent luminosity evolution of AGNs on the spectral index $\alpha$ as well as on the index $x$ in the integrated source counts is taken into account and one uses thus derived effective $\alpha$ and $x$ values 
in Eq.~(\ref{eq:1}) to derive the peculiar velocity, one gets results consistent with the CMB dipole, in accordance with the CP (Guandalin et al. 2022). Here, however, one has to be careful as the relevant values of $\alpha$ and $x$ to be  used in Eq.~(\ref{eq:1}), where $\cal D$ is estimated from  number counts, are not those derived for the whole AGN population, but the values at the threshold flux density of the sample being used. To understand that, one needs to consider the genesis of the dipole asymmetry in number counts due to a peculiar motion of the observer.
%-----------------------------------------------------------------
\begin{table*}
\caption{\label{T4}}{Peculiar velocity, $\bf v$, of the Solar system derived from various datasets using different techniques.}
\begin{tabular}{@{}ccccccccc}
\hline\hline
(1)&(2)&(3)&(4)&(5)&(6)&(7)\\
Dataset & Waveband  &  Technique Employed  & RA & Dec  & $v$ & Reference\\
&&& ($^\circ$)& ($^\circ$) & ($10^3$ km s$^{-1}$)&\\
\hline
CMB & Microwave& CMB temperature dipole  & $168$&$-7$  & 0.37 & Aghanim et al. (2020)\\
NVSS  & Radio (1.4 GHz) & Dipole in sky brightness& $153\pm 9$ & $1\pm 8$  &   $1.6\pm 0.4$ &  Singal (2011)\\
%NVSS & $142\pm 13$&$31\pm 10$  &   $1.0\pm 0.2$ &  \cite{Se22}\\
TGSS & Radio (150 MHz) & Source count dipole & $162\pm 9$&$3\pm 8$ & $3.8\pm 0.3$ &  Singal (2019a)\\
%TGSS & $155\pm 10$&$0\pm 8$ & $5.6\pm 0.3$ &  \cite{12}\\\\
DR12Q &Optical & Quasar redshift dipole  &  $166\pm 10$&$-12\pm 15$ & $-2.4\pm 0.3$ &  Singal (2019b)\\
MIRAGN  &Mid IR  & Source count dipole & $148\pm 19$&$23\pm 17$ &   $1.7\pm 0.2$ & Singal (2021)\\\\
QSOs &Mid IR  & Hubble plot dipole   & $179\pm 25$&$42\pm 25$ &  $8.1\pm 1.9$ & Singal (2022a)\\
SNe Ia   & Optical & Hubble plot dipole & $173\pm 12$&$10\pm 9$ &   $1.6\pm 0.5$ &  Singal (2022b)\\
WISE & Mid IR & Source density dipole & $142\pm 6$&$-5\pm 6$  &  $0.8\pm 0.1$ & Secrest et al. (2022)\\
%WISE & $150\pm 8\$&$2\pm 8$  &  $1.0\pm 0.1$ & \cite{Dam22}\\ 
VLASS & Radio (3 GHz) & Source count dipole & $189\pm 12$&$42\pm 22$ &  $1.6\pm 0.3$ &  present work\\
RACS & Radio (887.5 MHz) & Source count dipole  & $197\pm 13$&$-43\pm 23$ & $2.7\pm 0.3$ & present work\\\\
\hline
\end{tabular}
\end{table*}
%-----------------------------------------------------------------

As per CP, in the absence of a peculiar motion, the number counts in a sample above an {\em observed} flux-density threshold, say $S_0$,  will be the same, apart from the statistical fluctuations, in all directions, irrespective of any redshift distribution or luminosity evolution of the AGN population. However, due to a peculiar motion of the observer, as a result of Doppler boosting, the rest-frame flux-density threshold will be  $S_0/\delta^{1+\alpha}$. Since~the integral source counts of the extragalactic source population follow a power law, $N(>S)\propto S^{-x}$ (see, e.g., Fig.~\ref{F1}), the~number of sources observed at $S_0$ will be those at $S_0/\delta^{1+\alpha}$ in the rest frame, which will be higher by a factor $\delta^{x(1+\alpha)}$, where both  $\alpha$ and $x$ are the values in the vicinity of sample flux-density threshold $S_0$. Thus the observed number counts will show a dipole asymmetry because sources will move into the sample (or out of it for the opposite hemisphere) by a factor $\delta^{x(1+\alpha)}\approx 1+x(1+\alpha)(v/c)\cos\theta$, which has a dipole term ($\propto\cos\theta$), and for the CMB dipole with $v/c=1.23\times 10^{-3}$, it changes the flux density threshold by a small factor $1.0022\cos\theta$ for $x\approx 1$ and $\alpha\approx 0.8$. The relevant values of $x$ and $\alpha$ are determined empirically from the actual observations in the vicinity of observed threshold flux density $S_0$ (Singal 2019a; Tiwari 2019) and with statistical uncertainties $\stackrel{<}{_{\sim}} 15\%$, it will affect the derived peculiar velocity from  Eq.~(\ref{eq:1}) to be well within $\stackrel{<}{_{\sim}} 10\%$. Certainly, these will not bring the peculiar motion $p\sim 4$ close to $p\sim 1$, required for the CP to hold good, as claimed in Guandalin et al. (2022). 
Of course, due to the aberration of light, there is another factor $\propto \delta^{2}$ in the observed number densities, which enters Eq.~(\ref{eq:1}), but that, in any case being  independent of $\alpha$ or $x$, is not the point of contention here.

The results for various dipoles are summarized in Table~\ref{T4}, which is organized in the
following manner: 
(1) Dataset used.
(2) Waveband of observations. 
(3) Technique employed to compute the dipole.
(4) Estimated RA($^\circ$) of the corresponding dipole.
(5) Estimated Dec($^\circ$) of the corresponding dipole.
(6) Inferred peculiar velocity, $v$, in units of $10^3$ km s$^{-1}$.
(7) Reference to entries in columns (1) to (6).
From Table~\ref{T4}, we see that the peculiar velocity, derived from different datasets, using different techniques, varies by almost an order of magnitude. Combined together, there seems to be an almost overwhelming evidence that the peculiar velocity of the Solar system estimated from the distant radio source distributions in sky is not in concordance with what inferred from the CMB dipole anisotropy.

\begin{figure*}
\includegraphics[width=\linewidth]{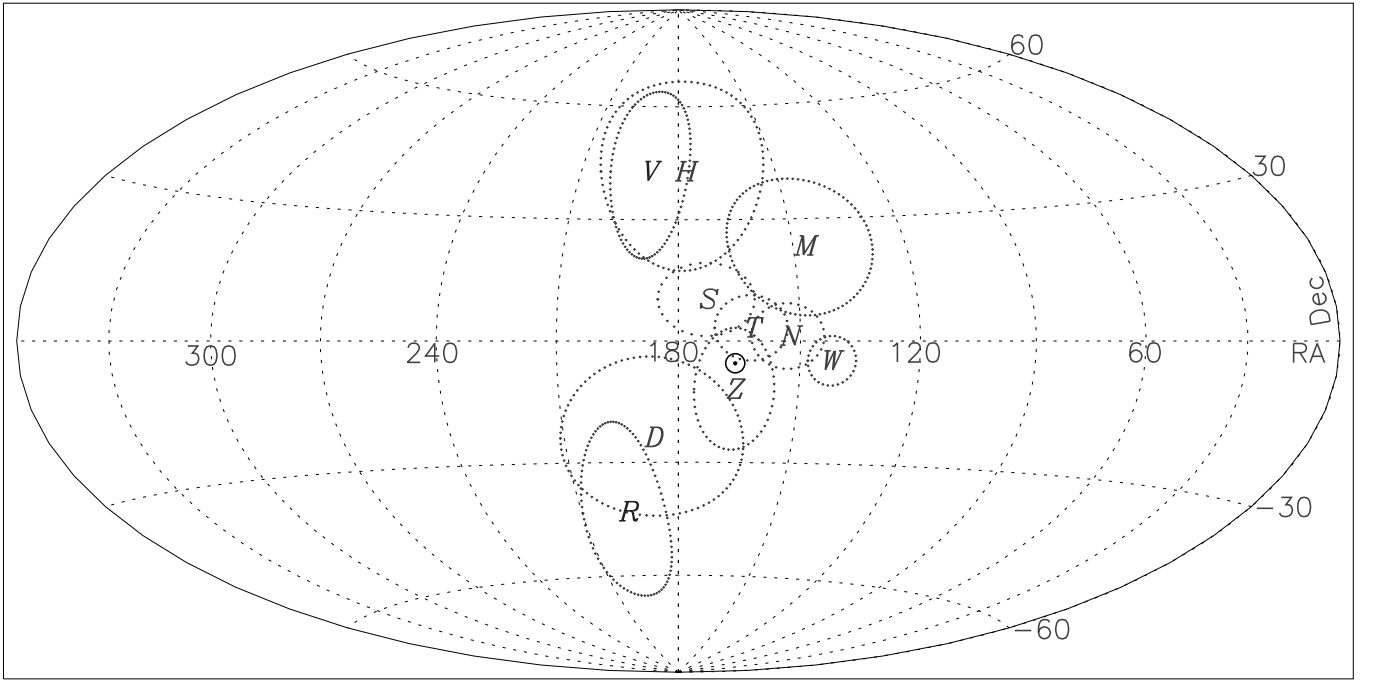}
\caption{The sky, in~the Hammer--Aitoff equal-area projection, plotted in equatorial coordinates RA and Dec, showing the positions of~the poles determined from the VLASS and RACS samples, indicated by $V$ and $R$ respectively, along with their error ellipses at $1\sigma$ ($68.3\%$) confidence limits. Also shown on the map are the other pole positions for various dipoles along with their error ellipses, $N$ (NVSS), $T$ (TGSS), $Z$ (DR12Q), $M$ (MIRAGN), $H$ (Hubble plot of quasars), $S$ (SNe Ia), 
%$N1$ (NVSS \cite{Se22}),  $W1$ (WISE \cite{Dam22}),
$W$ (WISE) and $D$ (Dark Flow). The~CMB pole, at RA$=168^{\circ}$, Dec$=-7^{\circ}$, indicated by $\odot$, has negligible errors. \label{F9}}
\end{figure*}

The sky positions of the poles determined from the VLASS and RACS samples, indicated by $V$ and $R$ respectively, are shown along with the error ellipses in Fig.~\ref{F9}. Also shown are the pole positions for other dipoles, along with their error ellipses:  $N$ (NVSS, Singal 2011), $T$ (TGSS, Singal 2019a), $Z$ (DR12Q, Singal 2019b), $M$ (MIRAGN, Singal 2021), $H$ (Hubble plot of quasars, Singal 2022a), $S$ (SNe Ia, Singal 2022b) and $W$ (WISE, Secrest et al. 2022).
% and $W1$ (WISE \cite{Dam22}). 
We have also plotted the~``dark flow'' dipole, indicated by $D$, which is a statistically significant dipole found at the position of galaxy clusters in filtered maps of the CMR temperature anisotropies (Kashlinsky et al. 2010). The~CMB pole, at RA$=168^{\circ}$, Dec$=-7^{\circ}$, indicated by $\odot$, has negligible errors (Lineweaver et al. 1996; Hinshaw et al. 2009; Aghanim et al. 2020). 
%---------------------------------------------------

%---------------------------------------------------

The observed fact that these discordant dipoles, resulting from many independent surveys carried out in different wavebands, are pointing along a narrow band in sky, indicates that these dipoles are somehow related and the cause of the dipoles may be common, otherwise they could have been pointing in random directions in sky. 
However, not even a single determination of the peculiar velocity from any AGN survey data has yielded a value that may be considered as close to $370$ km s$^{-1}$, determined from CMB dipole asymmetry. 
The lowest value for the dipole is at least twice the CMB value, and is obtained from the WISE data (Secrest et al. 2022), although a more recent Bayesian analysis of the number count dipole from the same data has yielded a value about 2.7 times the CMB dipole (Dam, Lewis \& Brewer 2022). Recently Wagenveld et al. (2023)  applied Bayesian estimators to determine the cosmic radio dipole from a combination of NVSS and RACS catalogues and estimated a dipole that aligns with the CMB dipole direction but with an amplitude three times larger, at $4.8\sigma$ level. 
%Almost all other dipoles are much larger in amplitude, with $p\approx 4$, corresponding to $v \approx 1600$ km s$^{-1}$, being the most common value. 
Thus not only all AGN dipoles are much larger than the CMB dipole, there is also a statistically significant disparity among themselves in their amplitudes, which does not support a peculiar motion of the Solar system to be that common cause, 
including that for the CMB dipole. 

In the CMB maps (Lineweaver et al. 1996; Hinshaw et al. 2009; Aghanim et al. 2020), the argument for the Solar system peculiar motion is that according to the CP, the CMB temperature distribution on large scale should be isotropic. The observed dipole (about one part in a thousand) there is attributed to observer's peculiar motion. However, from different amplitudes of various dipoles 
one cannot arrive at a single coherent picture of the Solar system peculiar velocity, which, defined as a motion relative to the local comoving coordinates, and from the CP a motion with respect to an average universe, should not depend upon the exact data or technique employed for its determination. Therefore such large discrepancies in the inferred velocity vectors may perhaps be a pointer to the need for some rethinking on the conventional interpretation of these dipoles. 

Could these dipoles, which exhibit some excess of source densities in certain sky directions, be due to some unaccounted for random fluctuations or some not-understood systematics in the data or the technique? 
%At least these dipoles, including the CMB dipole, do not seem to represent Solar system peculiar motion in sky as that should not vary with the data or the technique used to estimate it. 
Now in such a scenario the dipoles should be pointing in random directions in sky. However,  there are at least 9 statistically independent determinations of dipoles, counting MIRAGN and WISE samples as one point as they have partial overlap of data and thus may not be treated as completely independent data points. Also we could ignore the DR12Q point as the velocity direction is in opposite direction to that of CMB dipole. Further, as there are conflicting results for the SNe Ia, we could drop that point too. Even then there are seven dipole directions which seem to point in a narrow sky region ($\stackrel{<}{_{\sim}} 1/6$th of the total) about the CMB dipole,  which, by a conservative estimate, has a probability $\stackrel{<}{_{\sim}} (1/6)^7 < 10^{-5}$, to occur by a random chance.

Now various dipoles pointing toward approximately the same direction in the sky, if not due to observer's peculiar motion, seem to indicate an inherent preference for certain sky direction for these dipoles, which is not expected within the CP.
Therefore, one has to explain the pointing of all these dipoles in a narrow region of sky and what is so special about this direction in sky and whether it represents some sort of an ``axis'' of the universe. Further, various dipole magnitudes differing by as much as an order of magnitude, if interpreted as due to our peculiar velocity with respect to them, indicates that there may be a large relative motion of the various cosmic reference frames.
Significant differences in their derived peculiar velocities may be indicating that different cosmic reference frames are moving relative to each other or that the matter distribution on cosmic scales is not homogeneous and isotropic, in contravention of what expected from the CP, 
% All this does not fit with the cosmological principle, 
which is the starting point for the standard modern cosmology. There is other corroborating evidence that puts doubt on concurrency of the observable Universe with the CP (Aluri et al. 2023). Perhaps there is  need for a fresh look at the role of the CP in the cosmological models.

%--------------------------------------------
\section{Conclusions}\label{S5}
From the dipole anisotropies seen in the sky distribution of sources in the VLASS and RACS surveys, the inferred peculiar motion of the Solar system turns out a factor at least four times higher than what inferred from the CMB dipole. The directions of radio dipoles in the two datasets differ significantly not only from the CMB dipole but also from  each other. However these as well as other previously determined cosmic dipoles seem to point toward a relatively narrow region of the sky, which crudely speaking, has a low probability, less than one part in $10^{5}$, to occur by a random chance. From the different amplitudes of various dipoles we cannot arrive at a single coherent picture of the Solar system peculiar velocity, which, should not depend upon the exact data or technique employed for its determination. Therefore such large discrepancies in the inferred velocity vectors may perhaps be a pointer to the need for some rethinking on the conventional interpretation of these dipoles as due to the observer's peculiar motion. This indicates an inherent preference for certain sky direction for these dipoles, which seems discordant with the cosmological principle, the basis of modern cosmology.
%--------------------------------------------
\section*{Data Availability}
The VLASS data used in this article are available in VizieR Astronomical Server in the public domain at http://vizier.u-strasbg.fr/viz-bin/VizieR. The dataset is downloadable by selecting catalog: J/ApJS/255/30/comp. Another, independent version of the VLASS catalogue can be found in the electronic edition of the Astrophysical Journal in FITS format at https://iopscience.iop.org/article/10.3847/1538-4357/abf73b/meta. 
The RACS catalogue is available at https://doi.org/10.25919/8zyw-5w85 under Files as the file  RACS\_DR1\_Sources\_GalacticCut\_v2021\_08.xml.
%--------------------------------------------
\section*{Acknowledgements}
I thank Katherine Hale for help in getting the required RACS data.
%%%%%%%%%%%%%%%%%%%%%%%%%%%%%%%%%%%%%%%%%%
\section*{Declarations}
The author has no conflicts of interest/competing interests to declare that are relevant to the content of this article. No funds, grants, or other support of any kind was received from anywhere for this research.
%--------------------------------------------
%--------------------------------------------

\end{document}